\documentclass[12pt,preprint]{aastex}

\begin{document}

\title{Filaments and Ionized Gas in the Vicinity of 3C 244.1
\altaffilmark{1} }
 
\author{Carlos Feinstein }
\affil{ Observatorio Astron\'{o}mico, Paseo del Bosque, 1900 La Plata,
Argentina \newline Space Telescope Science Institute, 3700 San Martin
Drive, Baltimore, MD 21218}
\email{cfeinstein@fcaglp.edu.ar}

\author{F. Duccio Macchetto\altaffilmark{2}}
\affil{Space Telescope Science Institute, 3700 San Martin Drive,
Baltimore, MD 21218}

\author{ Andr\'{e} R. Martel} 
\affil{Department of Physics and Astronomy, Johns Hopkins University,
3400 N. Charles Street, Baltimore, MD 21218}

\author{William B. Sparks} 
\affil{Space Telescope Science Institute, 3700 San Martin Drive,
Baltimore, MD 21218}

\altaffiltext{1}{Based on observations made with the NASA/ESA {\em
Hubble Space Telescope}, which is operated by the Association of
Universities for Research in Astronomy, Inc, under NASA contract NAS
5-26555}

\altaffiltext{2}{On assignment from the Space Science Department of
ESA}
 
\begin{abstract}
   We present results of {\em Hubble Space Telescope} (HST)
observations of the radio galaxy 3C 244.1. The broadband F702W ({\em
R}) and F555W ({\em V}) images (WFPC2/PC) show an elliptical galaxy
and gaseous filaments and blobs surrounding it. In the narrow-band
ramp filter, dominated by [O~III]$\lambda5007$, these filaments
are bright and have the same morphology as the broad band images. To
the south, the filaments have a cone-shaped structure and the radio
jet is located at the center of this cone. To the north of the galaxy,
the structure is found near the nucleus of the galaxy within its
elliptical profile. From the photometry, the two brighter structures
seem to be extended narrow line emission regions (ENLRs). The
comparison with diagnostic line ratios shows that the observed
emission is consistent with interactions between the expanding
radio-jet and the local denser medium.
\end{abstract}

\keywords{AGN - Radio Galaxies - Jets}

\section{Introduction}

   The {\em Hubble Space Telescope} (HST) has been used to undertake a
systematic survey of extragalactic radio sources, The 3CR Snapshot
Survey (de Koff et al. 1996; Martel et al. 1998,1999 ; McCarthy et
al. 1997, de Vries et al. 1997) selected from the {\em Revised 
Third Cambridge Catalogue of
Radio Sources} (3CR) (Bennett 1962a,b; Spinrad et al. 1985). These
data, taken with the Wide Field Planetary Camera (WFPC2) using the
F702W, F555W and ramp narrow-band filters, allow us to investigate the
relationships between the radio and optical morphologies in a large
sample of powerful radio galaxies over redhshifts of $0<z<1.5$.

   Imaging of nearby Seyfert galaxies with HST, such as Mrk 3, Mrk 6,
Mrk 573, NGC 1068, NGC 2992 and NGC 4151, (Capetti et al. 1995a,
1995b, 1996, 1997; Winge et al. 1997; Allen et al. 1998 ,1999; Axon et
al. 1998) has shown an intimate connection between the radio structure
and the extended NLR (ENLR).  These studies show that the interaction
of the radio jet with the ISM is the main source of UV photons that
ionize the ENLR in Seyfert galaxies.  In previous work using HST data
of 3C 299, Feinstein et al. (1999) found that the the NE radio lobe
lies within a shell-like structure of the ENLR, suggesting a physical
connection between the jet and this ENLR.  Evidence of this
interaction comes from the values of the line ratio diagnostics and
the [O~II]$\lambda3727$/[O~III]$\lambda5007$ ratio, an
estimator of the change of the ionization parameter $U$, over the
ENLR.  We are extending this work by investigating whether this
scenario is also applicable to other powerful radio sources.  In this
paper, we discuss the radio galaxy 3C 244.1

   3C 244.1 is a FRII radio source with two lobes with angular
separation of 52$''$ at position angle (P.A.) of $168^\circ$.  Both
lobes have the same flux density (Fernini et al. (1997) $\log({\rm
L}_{5 {\rm GHz}})\approx22$ W/Hz). Fernini et al. (1997) also detected
the central component of the source that coincides with the optical
identification of Spinrad et al. (1985), showing that the radio lobes
are located asymmetrically with respect to this central compact
source.

   At optical wavelengths, the host galaxy of 3C 244.1 was shown to
have strong emission lines and a redshift of $z=0.428$ (Spinrad et al.
1985). From CCD images, McCarthy et al. (1995) note that the galaxy
lies in a cluster, and their [O~III] image shows a high surface
brightness fan of emission extending $\sim7''$ to the northwest,
with a position angle similar to the radio structure.

   In this paper we discuss the morphology of 3C 244.1 as shown by the
high spatial resolution images of WFPC2, and we investigate the
possible scenarios that give rise to the observed emission morphology.

\section{Observations and data analysis}
 
   The HST/WFPC2 observations of 3C 244.1 were taken as part of the
3CR imaging Snapshot Survey (PI: Sparks) which was conducted in Cycles
4 to 8, in the F555W and F702W broad bands, and narrow emission-line bands
(de Koff et al. 1996; Martel et al. 1998,1999; McCarthy et al. 1997,
etc).  The 3C 244.1 images were obtained with the WFPC2/PC in April
1994 and May 1995 for the F702W filter and in July 1996 for the F555W
filter. Narrow-band images with the WFPC2/WF2 and the FR680N ramp
filter were taken in November 1995.  Table 1 shows the observation
log.
 
   At a redshift of $z=0.428$, the galaxy is at a distance of 1.5 Gpc
with a projected linear scale of 5.1 kpc arcsec$^{-1}$ (assuming
$H_{0}$=65 km sec$^{-1}$ Mpc$^{-1}$ and $q_{o}=0.5$).  The
WFPC2/PC scale is $0''.0455$ pixel$^{-1}$ and for the WFPC2/WF2 the
scale is $0''.0996$ pixel$^{-1}$, and therefore, the physical scales
for the images are 232 pc pixel$^{-1}$ for the PC mode and 492 pc
pixel$^{-1}$ for the WF2 mode.
 
   The reduction procedure for the F702W filter data, which is close
to Cousins {\em R} filter, is fully discussed in Martel et al. (1999). This
reduction includes the standard WFPC2 pipeline processing followed by
cosmic-ray removal.  The data reduction was carried out using IRAF and
the STSDAS package.  At the redshift of 3C 244.1, the F702W filter
includes the H$\gamma$, [O~III]$\lambda4363$, He~II$\lambda4686$,
H$\beta$, [O~III]$\lambda\lambda$4959,5007 and [N~I]$\lambda5199$
emission lines. The F555W filter data, which is close to Johnson {\em
V}, was reduced in the same manner and then registered and added to
produce one final image with higher S/N. Due to the redshift, the
F555W filter includes the flux from the emission lines of
[Ne~V]$\lambda3426$, [O~II]$\lambda3727$, [Ne~III]$\lambda3869$,
H$\delta$ and H$\gamma$ (the latter considerably attenuated by the
filter response).  In what follows we will define F702W as {\em R},
and F555W as {\em V}.  The FR680N is a narrow-band ramp filter, which,
given the redshift of the galaxy and its position on the CCD, produces
an [O~III]$\lambda5007$ image. These images were reduced as described
above and the final image was re-scaled to the resolution of the
WFPC2/PC.
 
   The F702W and F555W filter data were flux-calibrated using values
for the inverse sensitivities (PHOTFLAM) of $1.965\times10^{-18}$ ergs
cm$^{-2}$ \AA$^{-1}$ DN$^{-1}$, $3.49141\times10^{-18}$ cm$^{-2}$
\AA$^{-1}$ DN$^{-1}$, and zero-points for the Vega system of
M$_{photzpt}=-21.85$ and M$_{photzpt}=-21.07$,
respectively. Non-rotated images were used for these flux calibrations
to minimize interpolation errors.  The FR680N filter was calibrated
using the ramp filter calculator (Biretta et al, 1996).  An
approximate coordinate frame for the WFPC2 data is provided by the
image header information, based on the HST guide stars.  The accuracy
for the coordinates is approximately $1''$ (Biretta et al. 1996).
 
   Images obtained with different filters were registered to a common
reference frame using cross-correlation of the brighter elliptical
structures.  The sky background in each image was determined through
the statistical analysis of a $3\sigma$ clipping of the average
background value in several regions in each of the images.

\section{Results}

   The optical broadband images show an elliptical galaxy with blobs
and filaments (Fig. 1 and 2). Both the {\em V} and {\em R} images show
these structures, and some are also present in the ramp image. We have
named the most prominent of these structures as A,B,C,D and E
(Fig. 3).  To compare with radio data, we superimposed a 8.4 GHz radio
map from the observations of Leahy (1999) in Fig. 1.  The total
extension of the radio structure is $51"$ (from lobe to lobe), larger
that the size of these HST images.

   The object named A is bright and extended, and is located $2''.9$
from the nucleus (P.A. $125^\circ$), and has the same shape in all
filters.  The flux in the narrow-filter implies a large emission of
[O~III]$\lambda5007$ emission. The structure named B, is located
$2''.6$ away from the nucleus at P.A. $150^\circ$, and is similar in
shape to structure A albeit fainter.  Both of these structures (A and
B) seem to be connected with the galaxy by very faint filaments.

   The object named C is located to the west at a separation of
$1''.4$ from the nucleus at P.A. $210^\circ$.  A filament appears to
connected it directly to the center of the galaxy. Superimposed over
this line at $1''$ from the nucleus, there is another blob (C'). The
filament just described and those associated with blobs A and B, can
be interpreted as the edges of a biconic structure with an opening
angle of $85^\circ$.  This kind of morphology has been found to be
common in a variety of active galaxies ranging from nearby spirals
(for example NGC 1068, Evans et al. 1991; NGC 5728, Wilson et
al. 1993) to powerful 3CR objects (e.g. 3C~299 Feinstein et al. 1999).
The southern radio-jet is located in the middle of this bi-conical
structure.

   The structure D is located inside the elliptical profile of the
galaxy, NE of the nucleus. After the subtraction of a $r^{1/4}$
profile (using the IRAF/STSDAS task {\em ellipse}) from the image, a
filamentary link emerges connecting D to the galaxy nucleus (see
Fig. 1.).  This structure seems not to be as large as the others and
is very difficult to measure in the broad-band images because of the
steep shape of the elliptical galaxy profile. This structure is also
evident in the [O~III]$\lambda5007$ image, but as for the broad-band
data, it is not possible to measure its flux reliably.

   The structure named E, at $3''.3$ from the nucleus at
P.A. $110^\circ$, is not present in the narrow-band filter. Its
profile is compact and steep, showing all the typical characteristics
of a field galaxy with a redshift different from that of 3C 244.1 .

   We have carried out photometry of the structures surrounding
3C~244.1 using the IRAF task {\em polyphot} (from the NOAO APPHOT
package). This IRAF task computes the magnitude of an object in an
image inside a polygonal aperture. The sky background is also computed
and subtracted and the errors are calculated.  To measure the bulk of
the emission for each structure a polygon that closely matches the
shape of the region was careful chosen.  The same polygons were used
on all images from the different filters to make the data comparable
between these filters. However, for the {\em V} filter, where the
signal is lower, some of these structures were difficult to measure
above the background noise.

   Table 2 shows the result of the photometry using the polygonal
measurement technique and the subsequent calibration.  For the
broad-bands filters, the data were calibrated using the inverse
sensitivity (PHOTFLAM) and the filter width (see Wide Field and
Planetary Camera Instrument Handbook - Biretta, 1996). For the FR680N
filter, the calibration was done using the IRAF/STSDAS {\em synphot},
and checked with the WFPC2 Exposure Time Calculator (ETC). The flux
observed at this filter, basically the flux of the
[O~III]$\lambda5007$ line, was corrected by a small contamination
($\sim12$\%) of the [O~III]$\lambda4959$ line (asumming that the flux
of the [O~III]$\lambda5007$ line is 2.88 times the
[O~III]$\lambda4959$ flux). This correction was calculated by
computing the central wavelength at the position on the CCD, and
assuming that the filter is Gaussian, centered at that wavelength with
a FWHM of 1.3\% (value obtained from the ETC) of this
central wavelength. Columns 2, 3 and 4 of Table 2 are the results of
these calibrations.

From Fig. 1, structures A and B are very close the jet flow, so the first 
attempt was to test if these objects were optical synchrotron knots on 
the radio jet.
So we fit a power law to flux from the broad band filters F555W and F702W.
If the main source of flux is synchrotron radiation, the flux must follow
a power law with an exponent in the range -0.5 to -1.5 .
To make the test as more accurately as we can, we used the task 
{\em calcphot-synphot} 
(from the SYNPHOT/STSDAS IRAF package) to integrate the power law under
 a better simulation of each filter. We found that the observed 
flux is not a power law and the best fit is a flat distribution with a null 
slope. 
So, we conclude that the observed flux is inconsistent with  power-law 
distributions of optical synchrotron radiation.

From the the broad band filters, we can estimated the (V-I) colors of 
the elliptical host galaxy, A and B regions. Colors of
the other structures can not be measured unless we have a better 
modeling of the underline galaxy flux, for which we need deeper observations 
and better spatial resolution. Objects A and B are far away
from the elliptical profile and are clearly not contaminated.
For the elliptical galaxy we obtain (V-I)=1.2 $\pm$0.3 which is close
to the value predicted by  R\"onnback et al. (1996) for an elliptical 
galaxy at the redshift
of 3C 244.1 (z=0.428) meanwhile the objects A and B, seems to be
bluer with values -0.2 $\pm$ 0.2 and -0.6 $\pm$ 0.5 respectively.  


Both structures A and B, given the large flux in the narrow ramp filter, are
obviously at the 3C 244.1 redshift. Also both show the same morphology shape
as the  broad-band images. Since the {\em R} filter data also
includes the flux of H$\gamma$, [O III]$\lambda4363$,
He~II$\lambda4686$, H$\beta$, [O III]$\lambda\lambda$4959,5007 and
[N~I]$\lambda5199$ emission lines, the total flux measured in this
broad band must be larger than the ramp filter which only includes
[O~III]$\lambda5007$. The fact that they are so similar can be
explained if we assume than the flux originates in a pure emission
line region, without any significant continuum contribution. 

   As the narrow-band filter data shows, there is a large amount of
flux in the emission line [O~III]$\lambda5007$. Thus it is very
reasonable to suggest that the structures A and B are Extended Narrow
Line Regions (ENLR). The ENLR have been found to be associated with a
wide variety of AGN, from Seyfert 2 to radio galaxies.

   Column 5 (Table 2) shows the flux calibration of the broad-band
{\em R} filter assuming that all the flux is from a source that has a
spectrum dominated by bright emission lines.  The shape of the F702W
filter is practically flat (See Wide Field and Planetary Camera 2
Instrument Handbook, page 217), and so it is easy to compute the
calibration of the flux of all the lines by calculating the total flux
for only one emission line (that accounts for all the line emission
flux).  To make this calculation, the ETC was used and the quantum
efficiencies (system + filter + CCD, hereafter QT) for each wavelength of
the conspicuous emission lines included in the {\em R} filter band
were computed.  Table 3 shows the results of this calculation and
confirm that all these emission lines are practically attenuated by
similar amounts. Therefore we used H$\beta$ as an equivalent line
(i.e. a line that has all the flux of all the lines included in the
filter) for the calibration. The error of this procedure is extremely
small.

\section{Discussion}

   A key question is to identify the mechanism responsible for the
ionization of these structures and filaments. The most likely
mechanisms are~: photoionization by the AGN, photoionization by the
AGN with matter-bounded clouds and shock-ionization by the radio
jet. We discuss each of these mechanism in detail.

\subsection{Photoionization by the AGN}

   This mechanism assumes that the UV flux produced by the AGN's
nuclear source photoionizes the line emission region.  The physical
status of the gas at any place can be described with the ionizing
parameter ($U$), which considers the dilution of the UV photons as the
distance from the nucleus increases.  For regions A and B, it is very
unlikely that the AGN is the only source of photons because the large
distances. To account for this scenario, some reference numbers of the
total photon budget can be calculated.

   Rawlings et al. (1989) performed spectroscopy of the nuclear zone
of 3C~244.1 and show a relation of [O III]$\lambda5007$
/H$\beta=4.5$. Using the results of Ferland and Netzer (1983), which
assume that the main source of photoionization continuum is a power
law ($f_{\nu} \propto \nu^{-1.5})$ and that the gas has a density of
$N_{H}=10^{3}$, they calculate several line emission ratios. Using
these results, it is straightforward to compute the ionization
parameter ($U$). This leads $U=3.2\times10^{-4}$ for solar abundances
and $U = 6.3\times10^{-4}$ for subsolar abundances.  This spectrum
covers only light coming from the nuclear zone that is observed in the
narrow band filter (regions A and B are not included in this
spectrum), so it is possible to calculate the total UV photon emission
of the nucleus. Because $Q=U4\pi r^2 N_{e} c$ and $r$ can be estimated
from the WFPC2 emission line image, this gives $Q=4.5\times10^{54}$
photons sec$^{-1}$ for solar abundances and $Q=9\times10^{54}$ photons
sec$^{-1}$ for sub-solar abundances. Due to the large distances, these
two values (for each this different abundances) are not enough to
ionize the blobs A and B. Therefore, we conclude that direct
photoionization by the nucleus is unlikely to be the dominant
mechanism responsible for the observed ENLR emission.

\subsection{Photoionization by the AGN with matter-bounded clouds}

   This idea was developed by Binette et al. (1996) in order to
explain the discrepant high ionization line ratios and high electron
temperatures observed in many active galaxies. These models are based
on a parameter $A_{M/I}$ which is the solid angle covered by the
``matter-bounded'' component relative to that covered by the
``ionization-bounded''component. In another paper, Binette et
al. (1997) refined their models.  Taking into account the possibility
of having a larger $U$ parameter, they derived three models with
values of $U=0.5$, 0.05, 0.02 and a density $n_{e}=1000$ cm$^{-3}$ for
the ``matter bounded'' clouds, which are exposed to ionization
radiation from the AGN.

   As we only have one line and the integrated total flux of several
lines from the {\em R} filter, we computed the flux ratio
$C_{D}=R$/[O III]$\lambda5007$ noting that the bulk of the {\rm R}
filter flux is the sum of H$\gamma$, [O~III]$\lambda4363$,
He~II$\lambda4686$, H$\beta$, [O~III]$\lambda\lambda$4959,5007 and
[N~I]$\lambda5199$ fluxes.  Note that the contribution from the
stellar host galaxy is not significant at the location of the blobs.
The line ratio can be derived from Table~2 and are $C_{D}=2.6$ for
structure A and $C_{D}=2.0$ for structure B.

   We computed $C_{D}$ (the ratio of lines include in the {\em R} filter to
[O~III]$\lambda5007$) for the results of mixed models of the papers
mentioned above.  Table~4 shows these results of $C_{D}$ as a function
of the $A_{M/I}$ parameter for the model of Binette et al. (1996) and
models H,M,L of (Binette et al. 1997).  Models M and L seem to fit the
observed $C_{D}$ for structures A and B, for $A_{M/I}\sim0.2$ and
0.34 for each object. But both of them have the same problem that we
have discuss for the first case of photoionization by the AGN,
$U=0.05$ for Model M and $U=0.02$ for model L, values that imply a
very high $Q$ ($Q=c n_{e} U d^{2}$) due to the distances to the galaxy
nucleus.  This calculation makes the mixed models very unlikely.


\subsection{Shocks}

   Fig 1 shows that the ENLR structures A and B are located close to
the edges of the radio-jet and it is possible that there is an
interaction between the jet and these regions.  It is important to
study whether there is a physical relationship between the radio-jet
and the optical line emission.  Taylor et al. (1992) proposed that
fast bowshocks resulting from the interaction of the radio jet and the
ISM were the source of ionizing photons of the emission-line gas in a
number of sources.  Capetti et al. (1995a, 1995b, 1996, 1997) and
Winge et al. (1997) were the first to show that this mechanism best
explains the optical emission in the NLR in nearby Seyfert galaxies
(Mrk 3, Mrk 6, Mrk 573, NGC 1068, NGC 4151 and NGC 7319).  Recently,
this work has been confirmed by Aoki et al. (1999) and Kukula et
al. (1999). In the case of the powerful radio galaxies of the 3CR
catalogue, Feinstein et al. (1999) showed that this interaction also
occurs in 3C~299, where clearly the NE radio-jet and the ENLR have
similar morphologies, and where there is further evidence of this
interaction from the values of the different emission-line ratios and
the evolution of the line ratio
[O~II]$\lambda3727$/[O~III]$\lambda5007$, as an estimator of the
changes of $U$, over the region.
 
   Dopita \& Sutherland (1995a,b) have modeled in detail the
ionization of the ENLR due to shocks. In one scenario, which has been
shown to work for Seyfert galaxies, the radio jet interacts with the
local interstellar medium and shocks the gas. In this scenario, the
hot post-shock plasma gas produces photons that can diffuse upstream
and downstream of the jet. Photons diffusing upstream can encounter
the preshocked gas and produce an extensive precursor HII region,
while those traveling downstream will influence the ionization and
temperature structure of the recombination of the shock.

   To check the validity of this interpretation, we use the line
diagnostic test of Dopita \& Sutherland (1995b).  We applied the same
procedure as the case above and use the flux ratio 
$C_{D}=R$/[O~III]$\lambda5007$.
On the other hand, we can compute the value of this ratio for
emission lines arising in a shocked ENLR.  Using Table 1 of Dopita \&
Sutherland (1995b) we calculated the total integrated flux for the
lines involved and the ratio to [O~III]$\lambda5007$ flux as~:

$ C_{D}= \frac {\sum_{i} F_{i} QT_{i}}  { [O III]\lambda 5007\ QT_{[O III]\lambda 5007} }$

where {\em i} represents each of the lines mentioned above and $F_{i}$
their respective flux.

   Table~5 shows the results of this computation, for shock speeds of
100, 150, 200 and 500 km sec$^{-1}$, with and without a precursor H~II
region. From this table, the values observed for structures A
($C_{D}=2.6$) and B ($C_{D}=2.0$) are easily explained with shock
models of $\sim200$~km sec$^{-1}$.  We conclude that shock
interactions between the expanding jet and the local denser medium are
responsible for the observed ENLR emission in 3C 244.1

\section{Conclusions}

   We have shown that radio galaxy 3C 244.1 has a filamentary
structure and some blobs bright in emission line
([O~III]$\lambda5007$). To the south, this structure seems to be an
emission-line cone with an opening angle of $85^\circ$ (from
P.A. $125^\circ$ to P.A $210^\circ$).  The radio-jet is located at the
center of this cone (P.A. $168^\circ$).  The two brighter blobs (named
A and B) are likely part of the ENLR (by comparing the flux between
broad-band and narrow-band), similar to those associated with
AGNs. These structures are larger and located far away from the
nucleus ($2''.9$). To the north more of these structures were found,
but near and associated with the nucleus.

   We tested several scenarios to explain the source of energy for the
emission lines observed and we found that the direct photoionization
by the AGN can only be possible if an unlikely large amount of UV
photons is provided by AGN. The same is true in more sophisticated
mixed medium models (Binette et al. 1996,1997).  The observations
seem to fit well the behavior of a model material
shocked by the radio jet (Dopita et al. 1996b). A shock velocity of
200 km sec$^{-1}$ can explain the ratio of the lines included in the
{\em R}-band filter to the [O~III]$\lambda5007$ line-emission.

   Therefore, from the morphology (location of the radio-jet) and
these physical arguments (the flux measured is consistent with
emission from shocked gas), we conclude that the ENLR structures of 3C
244.1 are the result of the interaction of the radio jet with ISM gas.

\acknowledgments
 
C.F. acknowledges the support from the STScI visitor program. We are
very grateful to P. Leahy for providing the radio map.


\begin{deluxetable}{c c c c}
\tablecolumns{4}
\tablewidth{0pc}
\tablecaption{Log of Observations}
\tablehead{\colhead{Filter Name} & \colhead{Emission Line} &
\colhead{Exp. Time} & \colhead{Date} \\
\colhead{\ \ } & \colhead{\ \ } & secs & \colhead{\ \ }}
\startdata
F702W   & Broad band\tablenotemark{1}           & 300 & 1994 Apr 17 \\
F702W   & Broad band\tablenotemark{1}           & 300 & 1995 May 28 \\
FR680N  & [O~III]$\lambda5007$ & 300 & 1995 Nov 13 \\
FR680N  & [O~III]$\lambda5007$ & 300 & 1995 Nov 13 \\
F555W   & Broad band\tablenotemark{1}           & 300 & 1996 Sep 7  \\
F555W   & Broad band\tablenotemark{1}           & 300 & 1996 Sep 7  \\
\enddata
\tablenotetext{1}{Due to the filter width, the measured flux includes emision from from several lines, see text }
\end{deluxetable}


\begin{deluxetable}{c r r r r}
\tablecolumns{5}
\tablewidth{0pc}
\tablecaption{Total Flux\tablenotemark{a}{\ \ }from the Structures}
\tablehead{\colhead{Region} & \colhead{F702W} & \colhead{F555W} 
& \colhead{[O~III]$\lambda5007$} & \colhead{F702W\tablenotemark{b}}}
\startdata
A & 17.26 $\pm$ 1.37 & 20.57 $\pm$ 2.69 & 9.36 $\pm$ 1.24 & 24.12 $\pm$ 1.91 \\
B &  5.25 $\pm$ 0.82 &  5.04 $\pm$ 1.62 & 3.58 $\pm$ 0.77 & 7.34  $\pm$ 1.14 \\
C &  2.6  $\pm$ 0.59 &   \nodata        &  \nodata        & 3.62  $\pm$ 0.83 \\
C'&  2.3  $\pm$ 0.52 &   \nodata        &  \nodata        & 3.24  $\pm$ 0.74 \\
\enddata
\tablenotetext{a}{Flux units are $10^{-16}$ ergs cm$^{-2}$ sec$^{-1}$.}
\tablenotetext{b}{The F702W filter flux is calibrated assuming that
the flux distribution is completely dominated by emission lines. See
text.}
\tablecomments{Measurements with values less than three times the
error are not reported.}
\end{deluxetable}


\begin{deluxetable}{c c}
\tablecolumns{2}
\tablewidth{0pc}
\tablecaption{Quantum efficiency (system + filter + CCD) for F702W at
the line wavelength}
\tablehead{\colhead{Line} & \colhead{QT}}
\startdata
H$\gamma$                    & 0.133 \\
$[{\rm O~III}]\lambda4363$   & 0.139 \\
$[{\rm He~II}]\lambda4686$   & 0.132 \\
H$\beta$                     & 0.124 \\
$[{\rm O~III}]\lambda4959$   & 0.113 \\
$[{\rm O~III}]\lambda5007$   & 0.112 \\
$[{\rm N~I}]\lambda5199$     & 0.092 \\
\enddata
\end{deluxetable}


\begin{deluxetable}{l c c c c}
\tablecolumns{5}
\tablewidth{0pc}
\tablecaption{Photoinization by an AGN with matter-bounded clouds}
\tablehead{\colhead{$A_{M/I}$} & \colhead{A\tablenotemark{a}} & \colhead{H} 
& \colhead{M} & \colhead{L} \\ 
\colhead{\ \ \ } & \colhead{$U=0.04$} & \colhead{$U=0.5$} 
& \colhead{$U=0.05$} & \colhead{$U=0.02$}}
\startdata
    0.04 &   1.79 &   1.83 &   4.95 &   6.77 \\
    0.14 &   1.74 &   1.76 &   2.99 &   3.19 \\
    0.24 &   1.71 &   1.71 &   2.46 &   2.50 \\
    0.34 &   1.68 &   1.67 &   2.21 &   2.21 \\
    0.44 &   1.66 &   1.65 &   2.06 &   2.05 \\
    0.54 &   1.65 &   1.63 &   1.97 &   1.95 \\
    0.64 &   1.64 &   1.61 &   1.90 &   1.88 \\
    0.74 &   1.63 &   1.60 &   1.85 &   1.83 \\
    0.84 &   1.62 &   1.59 &   1.81 &   1.79 \\
    0.94 &   1.61 &   1.58 &   1.78 &   1.76 \\
    1.04 &   1.60 &   1.57 &   1.76 &   1.73 \\
\enddata
\tablenotetext{a}{Model from Binette et al. (1996), no special name
assigned in the original paper. For other models of Binette et
al. (1997), names are as in the original paper.}
\end{deluxetable}


\begin{deluxetable}{l c}
\tablecolumns{2}
\tablewidth{0pc}
\tablecaption{Shock Models}
\tablehead{\colhead{Shock Velocity} & \colhead{$C_{D}$} \\
\colhead{km sec$^{-1}$} & \colhead{\ \ \ }}
\startdata
150             & 2.29 \\
200             & 2.67 \\
200 + precursor & 2.34 \\
300             & 3.34 \\
300 + precursor & 1.67 \\
500             & 2.71 \\
500 + precursor & 1.47 \\
\enddata
\end{deluxetable}

\newpage

\begin{figure}
\caption{3C 244.1, {\em R}-band image taken with WFPC2/F702W filter
(grey scale), the contours are from the radio map (Leahy, 1999)} 
\vspace*{0.25 truein}
\hspace*{ 0.25 truein} 
\epsscale{0.7}
\end{figure}

\begin{figure}
\caption{3C 244.1, top-left~: {\em R}-band image (F702W), top-right~:
{\em V}-band image (F555W), bottom-left : {\em R}-band image with the
elliptical profile substracted (note the plume of emission to the
northwest), bottom-right~: [O~III]$\lambda5007$ (re-scaled from the
WFPC2/WF to the resolution of the WFPC2/PC)}
\vspace*{0.25 truein}
\hspace*{0.25 truein} 
\epsscale{0.7} 
\end{figure}

\begin{figure}
\caption{3C 244.1, {\em R}-band image with the identifications of the
structures. The straight line indicates the direction of the
radio jet.}
\vspace*{0.25 truein}
\hspace*{0.25 truein} 
\epsscale{0.7} 
\end{figure} 
\end{document}